\documentclass{birkjour}

\usepackage{amsmath}
\usepackage{amssymb}
\usepackage{verbatim}   
\usepackage{color}    
\usepackage{mathrsfs,epsfig}  
\usepackage{fontenc}
\usepackage{textcomp}
\usepackage{fix-cm}
\usepackage{array}
\usepackage{enumerate}
\usepackage{latexsym, amsfonts, amssymb}
\usepackage{color}
\usepackage{comment}
\usepackage{subfigure}
\usepackage{accents}

\numberwithin{equation}{section}

\theoremstyle{definition}

\newtheorem{remark}[equation]{Remark}

\begin{document}
\fontsize{9}{11}\selectfont

\title[Hollomon's Power-law Columns]{Critical Buckling Loads of the Perfect\\ Hollomon's Power-law Columns}

\author{Dongming Wei}
\address{%
Department of Mathematics\\
University of New Orleans\\
New Orleans, LA, 70148, USA}
\email{dwei@uno.edu}

\author{Alejandro Sarria}
\address{%
Department of Mathematics\\
University of New Orleans\\
New Orleans, LA, 70148, USA}
\email{asarria1@uno.edu}

\author{Mohamed Elgindi}
\address{%
Texas A\& M University-Qatar, Qatar}
\email{mohamed.elgindi@qatar.tamu.edu}



\begin{abstract}
In this work, we present analytic formulas for calculating the critical buckling states of some plastic axial columns of constant cross-sections. The associated critical buckling loads are calculated by Euler-type analytic formulas and the associated deformed shapes are presented in terms of generalized trigonometric functions. The plasticity of the material is defined by the Hollomon's power-law equation. This is an extension of the Euler critical buckling loads of perfect elastic columns to perfect plastic columns. In particular, critical loads for perfect straight plastic columns with circular and rectangular cross-sections are calculated for a list of commonly used metals. Connections and comparisons to the classical result of the Euler-Engesser reduced-modulus loads are also presented.
\end{abstract}

\maketitle

\section{Introduction}
\label{sec:introduction}

In 1744, Euler introduced the analytic formula $P_{cr}=(\pi/L)^2EI$ for calculating the critical buckling load of a perfect straight column of elastica subject to an axial end-load with the pinned-pinned boundary conditions. In this formula, $E$ denotes the Young's modulus of the material, $I$ the area moment of inertia, and $L$ the length of the straight column. The Euler's formula is limited to applications for elastic columns. For a prior history to the theory of elastic stability of columns, see \cite{LG}. In 1859, Kirchhoff extended the theory to geometrically non-linear large deflections and provided a solution of the deflection curve in terms of elliptic integrals. In around 1889,  Engesser (\cite{Engesser1}) presented the tangent modulus critical buckling load formula $P_{cr}^{(T)}=(\pi/L)^2E_TI$, replacing the Young's modulus by the tangent modulus $E_T$ defined as the slope of the stress-strain curve at the start of the plastic strain range, or at the elastic yield point. In 1895, Jasinsky (\cite{Jasinski1}) noted a problem with Engesser's theory, which in 1898, Engesser himself corrected (\cite{Engesser2}) by modifying his tangent modulus theory (or double modulus theory), and introducing the Euler-Engesser reduced modulus load. Assuming that the neutral axis passes through the centroid, it  has a simple form: $P_{cr}^{(R)}=(\pi/L)^2E_rI$ for $ E_r=\frac{EI_1+E_TI_2}{I},$ and where $I_1$ and $I_2$ represent the area  moments of inertia relative to the compressed portion and the strain reversed portions of the cross section at the buckling state. For a detailed description of the reduced modulus load and the terms used here see,  e.g., \cite{Chen}. Around that time, various  attempts were made to extend Euler's result for various elastic and/or  inelastic columns. For example, Engesser (1891) and  Haringx (1948-49) each derived a Euler-type load formula by taking into account the influence of transverse shearing force. These formulae  were presented by Timoshenko and Gere (1961)  in \cite{TG}. For details, also see \cite{KardomateasandDancila}. 

For a modern account of the history of stability of columns, names of people who made important contributions to the buckling analysis of beam columns, and the associated bifurcation theory, see \cite {ZB}. In particular, the seminal work of J. W. Hutchinson \cite{H} provides technical and historical details of the theory of plastic buckling of columns under axial compression and the associated bifurcation theory. To this day, many studies of stability of various columns are still based on Euler's and Engesser's assumptions. Furthermore, several formulas of the Euler-type are available in the literature for nonlinear beam columns. Also, the development of numerical methods, such as the finite element and finite volume methods, along with the development of high-speed computers, have allowed researchers to arrive at better approximations of buckling solutions, specially for nonlinear columns.  
Among other numerical methods, the incremental methods are frequently used to linearize the nonlinear equations and the nonlinear solutions are approximated by the linear ones with refined linear steps. The Euler-type formulas can be used successfully at each linear substep for linear materials undergoing large deformations or even for nonlinear materials. One of the most popular methods is the Rik's method. However, the method is time consuming and computationally expensive. The unconverged solution is used to indicate that the column is unable to carry any more load and buckling has occurred. The load applied to the column in the last converged step previous to the unconverged solution is then considered as the critical buckling load of the column. However, like the Rik's method,  most of these methods are performed under the assumption that the column is not perfectly straight. In most nonlinear finite element buckling analysis, the standard procedure is to initiate a small deformation on the column and then incrementally increase the magnitude of the load until a yield criteria is reached for a new stress-strain relationship. At this point, buckling is said to have occurred. Therefore, these methods of nonlinear buckling analysis are, in essence, the regular stress analysis for imperfect columns and the solutions are fundamentally different from buckling of a perfect column. Euler's critical load formula states clearly that even perfect elastic columns buckle. For additional buckling analysis and various classical Euler-type formulas for critical loads of elastic and plastic columns, see (\cite{Bloom}, \cite{Chen},  \cite{Schafer}, \cite{Schafer2}, \cite{Schafer3},\cite{Timoshenko},\cite{Ugural},  \cite{Wang} ). Moreover, many steel-column design manuals are written by using Euler-type formulae as the basic design guides due to their simplicity and convenience (see e.g. \cite{Grey},  \cite{Moen}, \cite{Narayanan}). Therefore, obtaining analytic formulae for critical buckling loads of  inelastic columns remains very important.

In this work, we provide critical buckling load and mode formulae for some perfect columns of a large class of work-hardening metals subject to three sets of boundary conditions. The nonlinear material properties are given by the Hollomon's stress-strain equation \cite{Hollomon}
\begin{equation}
\label{eq:h}
\sigma=K|\epsilon|^{n-1}\epsilon
\end{equation}
where the material is assumed to be isotropic. In (\ref{eq:h}), $\sigma$ and $\epsilon$ are, respectively, the true stress and true strain defined in the classical theory of elasticity and plasticity,  $K$ is the bulk modulus, and $n$ the hardening index. 
The Hollomon's materials are also referred to as Ludwik power law materials and the true stress and true strain are referred to as Cauchy stress and logarithmic plastic strain, see e.g., \cite{Chakrabarty} and \cite{YLing}. 
The novel critical buckling loads derived in this paper can be used to help designers of work-hardening columns determine more precisely the criteria for stability of columns made of Hollomon materials. These results can be useful for a variety of applications in the design of oil and gas pipelines \cite{Mohr}, transportation vehicles; such as airplanes, automobiles, ships (\cite{Hall},\cite{Cunat}) and earthquake-proof infrastructure made of high-strength metal columns. 

One of the boundary conditions we will examine in this paper, are the pinned-pinned boundary conditions (see (\ref{eq:bc})i) below). In this case, our extension of the Euler critical load formula will be given by $P_{cr}(n)=\frac{2n\pi_n^2}{(n+1)L^{2n}}KI(n)$ where $I(n)=\int_Ay^{n+1}dA$ denotes the generalized area moment of inertia\footnote[1]{Satisfying $I(1)=I$ for $I$ as in Euler's formula.} and $\pi_n=2\int_0^{\frac{\pi}{2}}{\left(\text{cos}\theta\right)^{\frac{n-1}{n+1}}d\theta}>0$. Notice that this formula reduces to Euler's formula when $n=1$. Furthermore, the corresponding buckling mode will be presented in terms of a generalized trigonometric sine function to be introduced in Section 2. For deriving  the critical load formulae,  we will consider the boundary value problem for the fourth-order, nonlinear equation
\begin{equation}
\label{eq:mainprob}
\begin{split}
\left[\,\left|u^{\prime\prime}(x)\right|^{n-1}u^{\prime\prime}(x)\,\right]^{\prime\prime}+\frac{P}{KI(n)}u^{\prime\prime}(x)=0,\,\,\,\,\,\,\,\,\,\,x\in(0,L),\,\,\,\,n\in(0,1]
\end{split}
\end{equation}
with either pinned-pinned (PP), pinned-slide (PS) or slide-slide (SS) boundary conditions
\begin{equation}
\label{eq:bc}
\begin{cases}
u(0)=u(L)=u^{\prime\prime}(0)=u^{\prime\prime}(L)=0,\,\,&\text{(PP)},
\\
u(L)=u^{\prime}(0)=u^{\prime\prime}(L)=u^{\prime\prime\prime}(0)=0,\,\,&\text{(PS)},
\\
u(0)=u(L)=u^{\prime\prime\prime}(0)=u^{\prime\prime\prime}(L)=0,\,\,&\text{(SS)}.
\end{cases}
\end{equation}
In (\ref{eq:mainprob}), $u(x)$ denotes the transverse deflection from the vertical load x-axis, $P$ represents an end axial compressive load, and $I(n)=\int_A |y|^{n+1} dA,$ the constant generalized area moment of inertia relative to the cross-section $A(x)$, which remains perpendicular to the axis. We remark that equation (\ref{eq:mainprob}) is a special case of the more general model
\begin{equation}
\label{eq:prob}
\begin{split}
\left[KI(n,x)\left|u^{\prime\prime}(x)\right|^{n-1}u^{\prime\prime}(x)\right]^{\prime\prime}+P u^{\prime\prime}(x)=0,\,\,\,\,\,\,\,\,\,\,x\in(0,L),\,\,\,\,\,n\in(0,1]
\end{split}
\end{equation}
which allows for a space-dependent, generalized moment of inertia. The derivation of (\ref{eq:prob}) can be carried out
by the standard Euler's assumption that the cross section of a column remains perpendicular to the axis at the 
deformed equilibrium state. See for instance \cite{WL} and \cite{W} where, regarding the term $Pu''$ as a transverse load on the beam, a similar equation is derived.

For simplicity, let us introduce a new spatial variable $\bar x=\frac{x}{L}.$ Then, (\ref{eq:mainprob}) becomes
\begin{equation}
\label{eq:probscaled}
\begin{split}
\left[\left|u^{\prime\prime}(\bar x)\right|^{n-1}u^{\prime\prime}(\bar x)\right]^{\prime\prime}+\lambda_n u^{\prime\prime}(\bar x)=0,\,\,\,\,\,\,\,\,\,\,\bar x\in(0,1)
\end{split}
\end{equation}
with boundary conditions 
\begin{equation}
\label{eq:bc2}
\begin{cases}
u(0)=u(1)=u^{\prime\prime}(0)=u^{\prime\prime}(1)=0,\,\,&\text{(PP)},
\\
u(1)=u^{\prime}(0)=u^{\prime\prime}(1)=u^{\prime\prime\prime}(0)=0,\,\,&\text{(PS)},
\\
u(0)=u(1)=u^{\prime\prime\prime}(0)=u^{\prime\prime\prime}(1)=0,\,\,\,&\text{(SS)},
\end{cases}
\end{equation}
and where the constant $\lambda_n$ is given by
\begin{equation}
\label{eq:critical}
\begin{split}
\lambda_n=\frac{PL^{2n}}{KI(n)}.
\end{split}
\end{equation}
Consequently, the critical buckling loads, which we denote by $P_{cr}(n),$ are given by the formula
\begin{equation}
\label{eq:genload}
\begin{split}
P_{cr}(n)=\lambda_n\frac{KI(n)}{L^{2n}}
\end{split}
\end{equation}
The remainder of the paper is organized as follows. In section \ref{subsec:ppps}, we derive analytic formulas for convex solutions to (\ref{eq:probscaled}) as well as expressions for the critical loads (\ref{eq:genload}) under PP and PS boundary conditions. In section \ref{subsec:ss}, we construct a solution to (\ref{eq:probscaled}) with sign-changing curvature under the SS setting (\ref{eq:bc2})iii). Although the BVP (\ref{eq:probscaled})-(\ref{eq:bc2})iii) may be physically plausible for columns with large cross-sections supported by rollers fixed to the ends of the columns, it is mainly considered here to demonstrate how formulas for solutions satisfying other boundary conditions may be derived through a rather intuitive approach. In section \ref{sec:cp}, we introduce simple criteria that allows us to write an equation relating our critical loads with the corresponding Euler-Engesser's reduced modulus load. Even though the comparison will be done for the case of PP boundary conditions only, a similar argument extends to other cases. Finally, in section \ref{sec:ex} we compute critical loads for beams of several well-known materials satisfying the PP boundary conditions and with either circular or rectangular cross-sections. These results are then compared to the Euler-Engesser's reduced modulus loads with the strain yield value $\epsilon_y=0.02$. \footnote[1]{Although this value may vary from material to material, we will not compare the results individually since they are not readily available for the materials considered here. Regardless, the actual comparison can be done once the yield is given by using the formula derived in section \ref{sec:cp}.} Also, we use \textsc{Mathematica} to numerically calculate the critical buckling shape, which, as we will see, agrees with our analytic result.


\section{Buckling of the Perfect Plastic Beams of Hollomon Work-Hardening}
\label{sec:intro}

\subsection{pinned-pinned and pinned-slide Boundary conditions}
\label{subsec:ppps}
In this section, we derive analytic formulas for (\ref{eq:genload}), and related quantities, corresponding to equation (\ref{eq:probscaled}) with either PP or PS boundary conditions. More particularly, we will examine the case of convex solutions, $u^{\prime\prime}(\bar x)\geq0$ on $[0,1].$ As a result, equation (\ref{eq:probscaled}) becomes
\begin{equation}
\label{eq:probfinal}
\begin{split}
\left[(u''(\bar x))^n\right]^{\prime\prime}+\lambda_n u^{\prime\prime}(\bar x)=0,\,\,\,\,\,\,\,\,\,\,\bar x\in(0,1)
\end{split}
\end{equation}
for $n\in(0,1]$ and $\lambda_n$ as in (\ref{eq:critical}). Furthermore, using the substitutions $w(\bar x)=u^{\prime\prime}(\bar x)$ and $v(\bar x)=w(\bar x)^n,$ reduces $(\ref{eq:probfinal})$ to
\begin{equation}
\label{eq:secondprob}
\begin{split}
&v^{\prime\prime}(\bar x)+\lambda_n v(\bar x)^{\frac{1}{n}}=0,\,\,\,\,\,\,\bar x\in(0,1)
\end{split}
\end{equation}
with boundary conditions 
\begin{equation}
\label{eq:bc3}
v(0)=v(1)=0,\,\,\,\,\,\,\,\,\,\text{(PP)},\,\,\,\,\,\,\,\,\,\,\,\,\,\,\,\,\,\,\,\,\,\,v^\prime(0)=v(1)=0,\,\,\,\,\,\,\text{(PS)}.
\end{equation}

\subsubsection{pinned-pinned boundary conditions}
\label{subsec:pp}
Consider (\ref{eq:secondprob}) with PP boundary conditions (\ref{eq:bc3})i). In \cite{Buttazzo1}, the authors studied the more general BVP:  
\begin{equation}
\label{eq:butt}
\begin{split}
&\left[\lvert v^\prime(\bar x)\rvert^{p-2}v^\prime(\bar x)\right]^{\prime}+\lambda v(\bar x)^{q-1}=0,\,\,\,\,\,\,\,\,\,\,\,\,v\geq0,\,\,\,\,\,\,\,\bar x\in(0,1),
\\
&v(0)=v(1)=0,
\end{split}
\end{equation}
for $p\in(1,+\infty)$ and $q\in[1,+\infty).$ They constructed the following implicit solution symmetric about $\bar x=1/2,$
\begin{equation}
\label{eq:sol1}
\begin{split}
\int_0^{v(\bar x)}{(1-s^q)^{-\frac{1}{p}}ds}=\frac{2\bar x}{q}B(1/q,1/p^\prime),\,\,\,\,\,\,\,\,\,\,&\bar x\in[0,1/2],
\\
v(\bar x)=v(1-\bar x),\,\,\,\,\,\,\,\,\,\,\,\,\,\,\,\,\,\,\,\,\,\,\,\,\,\,\,\,\,\,\,\,\,\,\,\,\,\,\,\,\,\,\,\,\,\,\,\,\,\,\,\,\,\,\,\,\,\,\,\,\,\,\,\,&\bar x\in[1/2,1]
\end{split}
\end{equation}
where $p^\prime\in(1,+\infty)$ is such that $1/p+1/p^\prime=1$ and $B(a,b)$ denotes the standard Euler-Beta function defined by
\begin{equation}
\label{eq:defbeta}
\begin{split}
B(a,b)\equiv\int_0^1{t^{a-1}(1-t)^{b-1}dt}=2\int_0^{\frac{\pi}{2}}{\left(\text{cos}\theta\right)^{2a-1}\left(\text{sin}\theta\right)^{2b-1}d\theta},\,\,\,\,\,\,\,\,a,b>0.
\end{split}
\end{equation}
Notice that (\ref{eq:secondprob})-(\ref{eq:bc3})i) and (\ref{eq:butt}) coincide when $p=2$ and $q=1+\frac{1}{n}$ for, say, $n\in(0,1].$ Now, let us introduce the generalized inverse sine function (\cite{Drabek}, \cite{Franzina1}):
\begin{equation}
\label{eq:geninvsine}
\begin{split}
\text{arcsin}_{n}(y)=\frac{q}{2}\int_0^{\frac{2y}{q}}{\frac{ds}{(1-s^q)^{\frac{1}{2}}}},\,\,\,\,\,\,\,\,\,\,\,\,\,\,y\in\left[0,\frac{q}{2}\right]
\end{split}
\end{equation}
and the generalized $pi$ constant 
\begin{equation}
\label{eq:genpie}
\begin{split}
\pi_{n}=2\,\text{arcsin}_{n}\left(\frac{q}{2}\right)=B\left(\frac{1}{q},\frac{1}{2}\right)=2\int_0^{\frac{\pi}{2}}{\left(\text{cos}\theta\right)^{\frac{n-1}{n+1}}d\theta},
\end{split}
\end{equation}
which satisfies $\pi_1=\pi.$ 
Amongst some of its properties, (\ref{eq:geninvsine}) defines a strictly increasing function of $[0,q/2]$ onto $\left[0,\pi_{n}/2\right]$ where the inverse of (\ref{eq:geninvsine}), which we denote by $\text{sin}_{n},$ is extended to $[-\pi_{n},\pi_{n}]$ by setting $\text{sin}_{n}\theta=\text{sin}_{n}(\pi_{n}-\theta)$ for all $\theta\in(\pi_{n}/2,\pi_{n}]$ and $\text{sin}_{n}\theta=-\sin_{n}(-\theta)$ for $\theta\in[-\pi_{n},0).$ By periodicity, we can then extend to all of $\mathbb{R}.$
As a result, setting $y=\frac{q}{2}v=\frac{n+1}{2n}v,\, n\in(0,1]$ in (\ref{eq:geninvsine}) and using (\ref{eq:sol1}), we find that the first eigenfunction of (\ref{eq:secondprob}), $v_n(\bar x),$ is given by
\begin{equation}
\label{eq:eigf1}
\begin{split}
v_{n}(\bar x)=\frac{2n}{n+1}\sin_{n}\left(\pi_n\bar x\right),\,\,\,\,\,\,\,\,\,\,\,\,\,\,\,\,\,\,\,\,\,\,\,\,&\bar x\in[0,1/2],
\\
v_{n}(\bar x)=\frac{2n}{n+1}\sin_{n}\left(\pi_n(1-\bar x)\right),\,\,\,\,\,\,\,\,\,\,\,&\bar x\in[1/2,1].
\end{split}
\end{equation}
Notice that for $n=1,\, v_{1}(\bar x)=\sin(\pi\bar x),$ which is simply Euler's classical buckling shape for elastic columns. See Figure \ref{fig:pinpinload} for a plot of $(\ref{eq:eigf1})$ for several $n\in(0,1].$ Next, we derive formulas for $u$ and the first eigenvalue $\lambda_{n},$ as well as the critical loads $P_{cr}(n).$ Setting $y=\sin_n(t)$ in (\ref{eq:geninvsine}) gives
\begin{equation}
\label{eq:der1}
\begin{split}
t=\frac{q}{2}\int_0^{\frac{2}{q}\sin_n(t)}{\frac{ds}{(1-s^q)^{\frac{1}{2}}}}
\end{split}
\end{equation}
for $q=1+1/n.$ Define
\begin{equation}
\label{eq:id0}
\begin{split}
\cos_n(t)=\left(1-\left(\frac{2\sin_n(t)}{q}\right)^q\right)^{\frac{1}{2}},\,\,\,\,\,\,\,\,\,\,\,\,t\in[0,\pi_n/2],
\end{split}
\end{equation}


so that differentiation of (\ref{eq:der1}) yields
\begin{equation}
\label{eq:dergensin}
\begin{split}
\frac{d}{dt}\sin_n(t)=\cos_n(t).
\end{split}
\end{equation}
Furthermore, (\ref{eq:id0}) provides us with the identity 
\begin{equation}
\label{eq:id}
\begin{split}
\cos_n^2(t)+\left(\frac{2\sin_n(t)}{q}\right)^{q}=1.
\end{split}
\end{equation}
The derivative of $\cos_n(\cdot)$ can then be obtained by differentiating (\ref{eq:id}) and using (\ref{eq:dergensin}). We obtain
\begin{equation}
\label{eq:dergencos}
\begin{split}
\frac{d}{dt}\cos_n(t)=-\left(\frac{2\sin_n(t)}{q}\right)^{q-1}.
\end{split}
\end{equation}
Now, $u^{\prime\prime}=v^{1/n}$ and (\ref{eq:eigf1}) imply
\begin{equation}
\label{eq:udouble}
\begin{split}
u_{n}^{\prime\prime}(\bar x)&=\left(\frac{2n}{n+1}\sin_n(\pi_n\bar x)\right)^{\frac{1}{n}},\,\,\,\,\,\,\,\,\,\,\,\,\,\,\,\,\,\,\,\,\,\bar x\in[0,1/2],
\\
u_{n}^{\prime\prime}(\bar x)&=\left(\frac{2n}{n+1}\sin_n(\pi_n(1-\bar x))\right)^{\frac{1}{n}},\,\,\,\,\,\,\,\bar x\in[1/2,1]
\end{split}
\end{equation}
for $\pi_n$ as defined in (\ref{eq:genpie}). Then using (\ref{eq:dergencos}), we deduce that
\begin{equation}
\label{eq:id4}
\begin{split}
u_{n}^{\prime}(\bar x)&=-\frac{1}{\pi_n}\cos_n(\pi_n\bar x)+C_1,\,\,\,\,\,\,\,\,\,\,\,\,\,\,\,\,\,\,\,\,\bar x\in[0,1/2],
\\
u_{n}^{\prime}(\bar x)&=\frac{1}{\pi_n}\cos_n(\pi_n(1-\bar x))+C_1,\,\,\,\,\,\,\,\,\,\,\bar x\in[1/2,1],
\end{split}
\end{equation}
and so, by formula (\ref{eq:dergensin}), the boundary conditions $u(0)=u(1)=0$ and evaluation at $\bar x=1/2,$ we find
\begin{equation}
\label{eq:first}
\begin{split}
u_{n}(\bar x)&=-\frac{1}{\pi_n^2}\sin_n(\pi_n\bar x),\,\,\,\,\,\,\,\,\,\,\,\,\,\,\,\,\,\,\,\,\,\,\,\bar x\in[0,1/2],
\\
u_{n}(\bar x)&=-\frac{1}{\pi_n^2}\sin_n(\pi_n(1-\bar x)),\,\,\,\,\,\,\,\,\,\,\bar x\in[1/2,1].
\end{split}
\end{equation}
Plots for $u_n$ may be generated parametrically from (\ref{eq:geninvsine}), (\ref{eq:first}) and the symmetry of the generalized function. Also, $\lambda_{n}$ can be obtained by substituting (\ref{eq:udouble}) into equation (\ref{eq:probfinal}) and using the identities (\ref{eq:dergensin}) and (\ref{eq:dergencos}). This yields
\begin{equation}
\label{eq:firsteig1}
\begin{split}
\lambda_{n}=\frac{2n\pi_n^2}{n+1},
\end{split}
\end{equation}
which satisfies $\lambda_{1}=\pi^2.$ Finally, by (\ref{eq:genload}) and (\ref{eq:firsteig1}),
\begin{equation}
\label{eq:critload1}
\begin{split}
P_{cr}(n)=\frac{2n\pi_n^2}{(n+1)L^{2n}}KI(n).
\end{split}
\end{equation}
\begin{remark}
\label{rem:rempp}
The critical load (\ref{eq:critload1}) grows unbounded as $n\to0^+.$ In fact, using the identity $\Gamma(x)\Gamma(y)=\Gamma(x+y)B(x,y),$ where $\Gamma(\cdot)$ denotes the standard gamma function (see appendix for definition), we can rewrite the generalized $pi$ constant (\ref{eq:genpie}) as $$\pi_n=\frac{\sqrt{\pi}\,\Gamma\left(\frac{n}{1+n}\right)}{\Gamma\left(\frac{3}{2}-\frac{1}{1+n}\right)},\,\,\,\,\,\,\,\,\,\,\,n\in(0,1].$$ This implies that $\pi_n\sim\Gamma(n)$ for $n>0$ small. But $\Gamma(y+1)=y\,\Gamma(y)$ for $y\in\mathbb{R}^+.$ Then, assuming a finite value for $I(n)$ as $n\to0^+,$ (\ref{eq:critload1}) implies that $P_{cr}(n)\sim\frac{C}{n}$ for $n>0$ small and some positive constant $C.$ Therefore, a smaller $n$ corresponds to a larger $P_{cr}(n)$, which is consistent with hardening.
\end{remark}

\subsubsection{pinned-slide boundary conditions}
\label{subsec:ps}

\begin{center}
\begin{figure}[!ht]
\centering
\subfigure{
\includegraphics[scale=0.35]{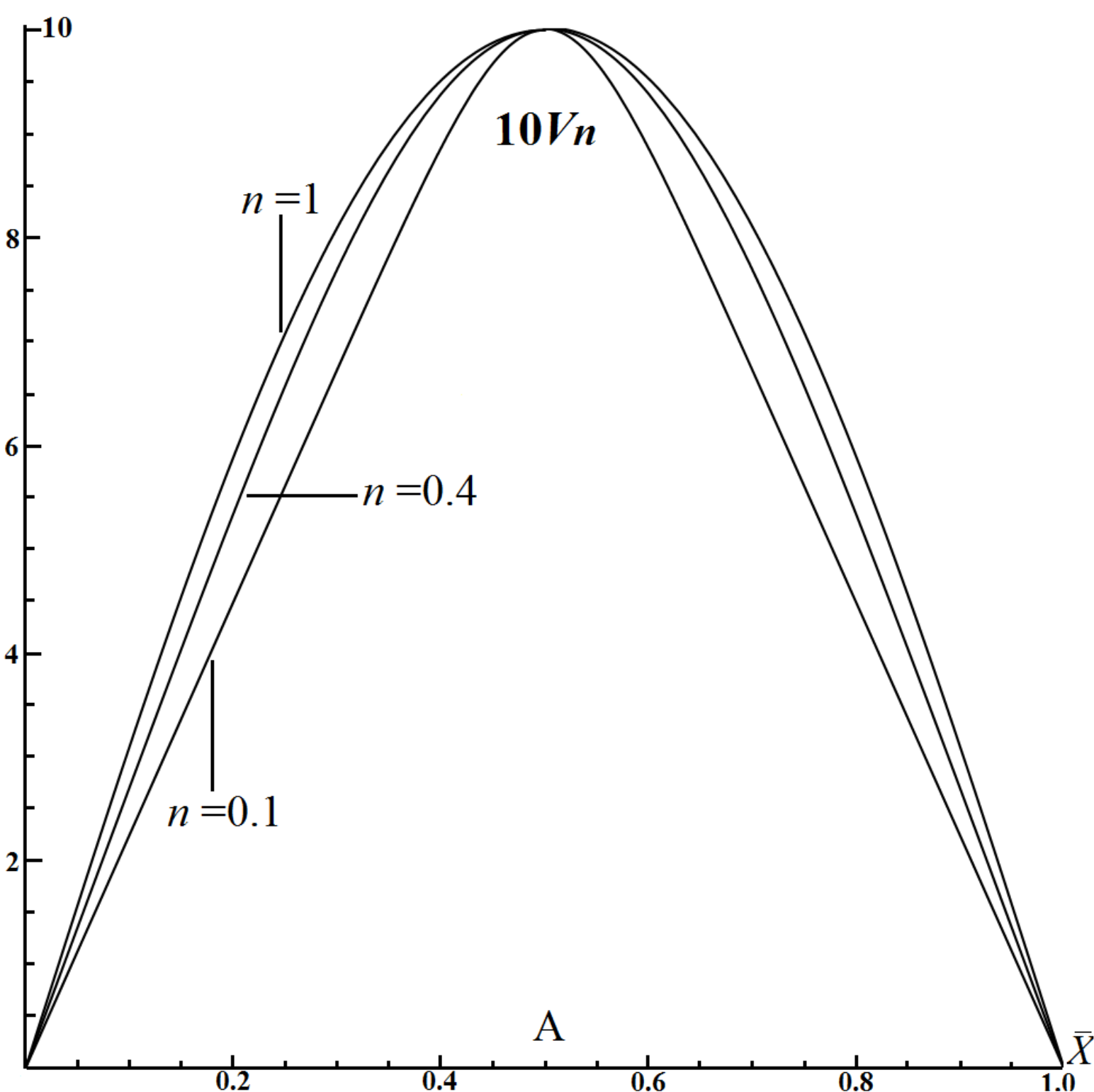}
\label{fig:pinpinload}
}
\subfigure{
\includegraphics[scale=0.35]{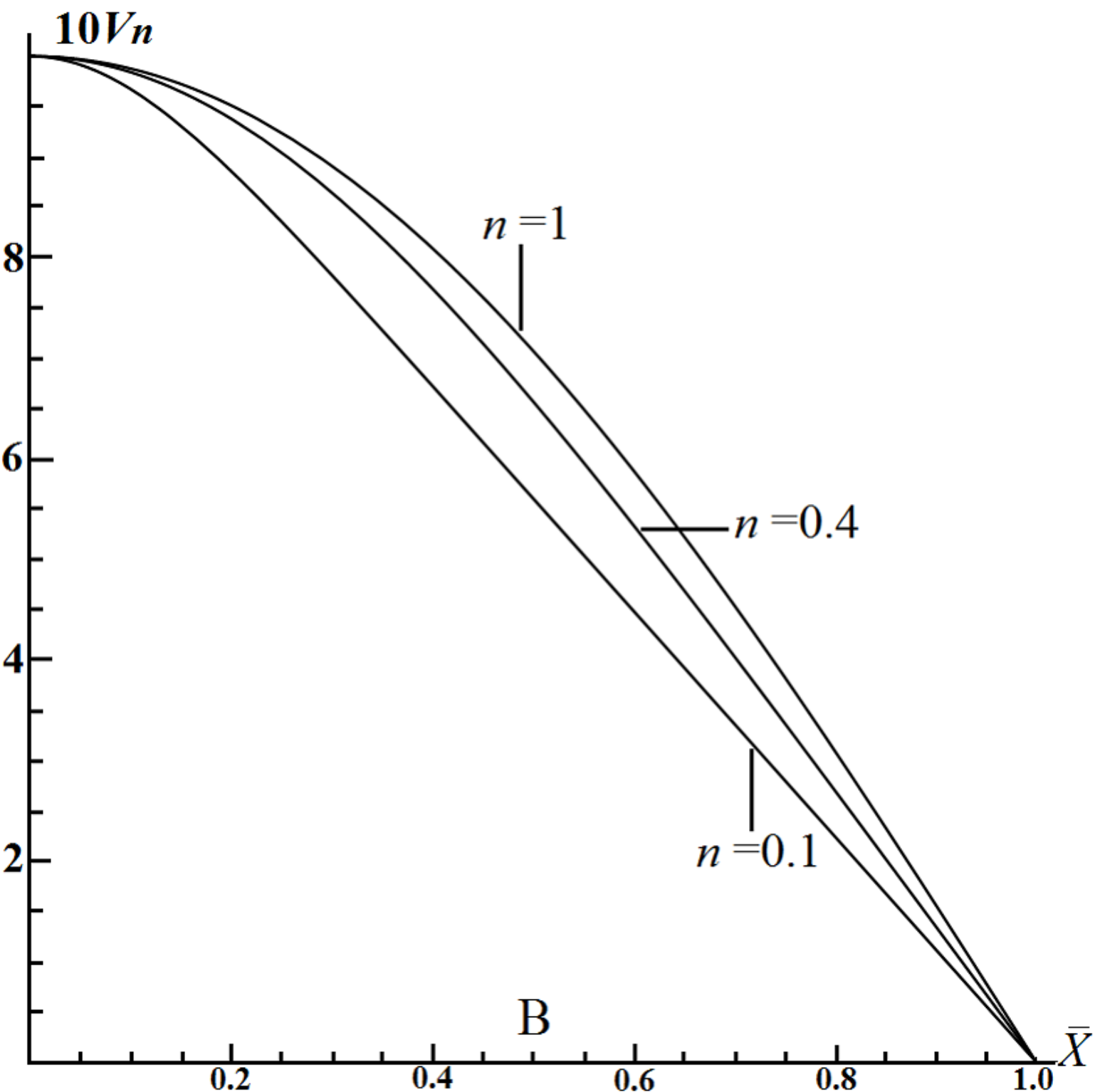}
\label{fig:pinslideload}
}
\label{fig:twofigs2}
\caption{Figures $A$ and $B$ depict, respectively, the eigenfunctions (\ref{eq:eigf1}) and (\ref{eq:eq9}) for $n=1,0.4,0.1$ and, in the case of (\ref{eq:eq9}), $\alpha=1$.}
\end{figure}
\end{center}

Next, consider (\ref{eq:secondprob}) with the PS boundary conditions $(\ref{eq:bc3})$ii).
Recall that under PP boundary conditions $u^{\prime\prime}(0)=0,$ whereas $u''(0)$ is not prescribed in the PS case. As a result, let us set $u^{\prime\prime}(0)=\alpha$ for some $\alpha\in\mathbb{R}^+.$ This condition is then normalized by setting $h(\bar x)=\alpha^{-n}v(\bar x),$ i.e. $h(0)=1$ (recall that $v=(u^{\prime\prime})^n$). Then, by the change of variables, BVP (\ref{eq:secondprob}) now reads
\begin{equation}
\label{eq:secondprob22}
\begin{split}
&h^{\prime\prime}(\bar x)+\lambda_{n}\alpha^{1-n}h(\bar x)^{\frac{1}{n}}=0,\,\,\,\,\,\,\,\,\,\,\bar x\in(0,1),
\\
&h^\prime(0)=h(1)=0.
\end{split}
\end{equation}
Multiplying (\ref{eq:secondprob22})i) by $h^\prime$ and integrating yields $$\frac{(h^\prime)^2}{2}+\frac{\lambda_{n}\,\alpha^{1-n}\,h^q}{q}=\frac{\lambda_{n}\,\alpha^{1-n}}{q}$$
for $q=1+1/n.$ In \cite{Buttazzo1}, it was shown (see the appendix) that $h^\prime$ is a strictly decreasing function on $(0,1).$ Then, since $h^\prime(0)=0$ and we only consider nontrivial solutions, this means $h^\prime<0.$ Consequently, the last equation implies $h^\prime(1-h^q)^{-\frac{1}{2}}=-\left(\frac{2\lambda_{n}}{q\alpha^{n-1}}\right)^{\frac{1}{2}},$ or equivalently
\begin{equation}
\label{eq:eq3}
\begin{split}
\int_0^{h(\bar x)}{(1-s^q)^{-\frac{1}{2}}ds}=-\left(\frac{2\lambda_{n}}{q\alpha^{n-1}}\right)^{\frac{1}{2}}\bar x+C_1.
\end{split}
\end{equation}
However $h(1)=0,$ then (\ref{eq:eq3}) gives
\begin{equation}
\label{eq:eq4}
\begin{split}
\int_0^{h(\bar x)}{(1-s^q)^{-\frac{1}{2}}ds}=\left(\frac{2\lambda_{n}}{q\alpha^{n-1}}\right)^{\frac{1}{2}}(1-\bar x),\,\,\,\,\,\,\,\,\,\,\,\,\bar x\in(0,1).
\end{split}
\end{equation}
As a result, using $(\ref{eq:geninvsine})$ and (\ref{eq:eq4}) yields
\begin{equation}
\label{eq:eq5}
\begin{split}
\frac{2}{q}\arcsin_n\left(\frac{q}{2}h(\bar x)\right)=\left(\frac{2\lambda_{n}}{q\alpha^{n-1}}\right)^{\frac{1}{2}}(1-\bar x),\,\,\,\,\,\,\,\,\,\,\,\,\,\,\,h\in[0,1].
\end{split}
\end{equation}
Now, setting $\bar x=0$ in $(\ref{eq:eq4})$ and using $h(0)=1,$ we obtain $\int_0^{1}{(1-s^q)^{-\frac{1}{2}}ds}=\left(\frac{2\lambda_{n}}{q\alpha^{n-1}}\right)^{\frac{1}{2}},$ so that, by $(\ref{eq:geninvsine})$ and $(\ref{eq:genpie}),$
\begin{equation}
\label{eq:firsteig2}
\begin{split}
\lambda_{n}=\frac{n\pi_n^2}{2(n+1)\alpha^{1-n}}.
\end{split}
\end{equation}
Finally, $(\ref{eq:critical})$ and $(\ref{eq:firsteig2})$ imply
\begin{equation}
\label{eq:critload2}
\begin{split}
P_{cr}(n)=\frac{n\pi_n^2}{2(n+1)\alpha^{1-n}L^{2n}}KI(n),
\end{split}
\end{equation}
and the eigenfunction $v_{n}$ is obtained from (\ref{eq:eq5}), (\ref{eq:firsteig2}) and $h(\bar x)=v(\bar x)\alpha^{-n}$ as
\begin{equation}
\label{eq:eq9}
\begin{split}
v_{n}(\bar x)=\frac{2n\alpha^{n}}{n+1}\sin_n\left(\frac{\pi_n(1-\bar x)}{2}\right).
\end{split}
\end{equation}
Notice that $v_{1}(\bar x)=\alpha\sin\left(\frac{\pi(1-\bar x)}{2}\right)=\alpha\cos\left(\frac{\pi\bar x}{2}\right).$ See Figure \ref{fig:pinslideload} for a plot of (\ref{eq:eq9}) with $\alpha=u^{\prime\prime}(0)=1.$ Finally, by following the argument in the PP case, we use $u^{\prime\prime}=v^{\frac{1}{n}},$ (\ref{eq:dergencos}) and (\ref{eq:eq9}), to find
\begin{equation}
\label{eq:sec}
\begin{split}
u_{n}^{\prime}(\bar x)=\frac{2\alpha}{\pi_n}\cos_n\left(\frac{\pi_n(1-\bar x)}{2}\right),
\end{split}
\end{equation}
where the constant of integration is zero due to $u^\prime(0)=0.$ Then, (\ref{eq:dergensin}), (\ref{eq:sec}) and $u(1)=0$ yield
\begin{equation}
\label{eq:sec2}
\begin{split}
u_{n}(\bar x)=-\frac{4\alpha}{\pi_n^2}\sin_n\left(\frac{\pi_n(1-\bar x)}{2}\right),\,\,\,\,\,\,\,\,\,\,\mathbb{R}^+\ni\alpha=u^{\prime\prime}(0).
\end{split}
\end{equation}


\subsection{Slide-slide boundary conditions}
\label{subsec:ss}
\begin{center}
\begin{figure}[!ht]
\centering
\includegraphics[scale=0.42]{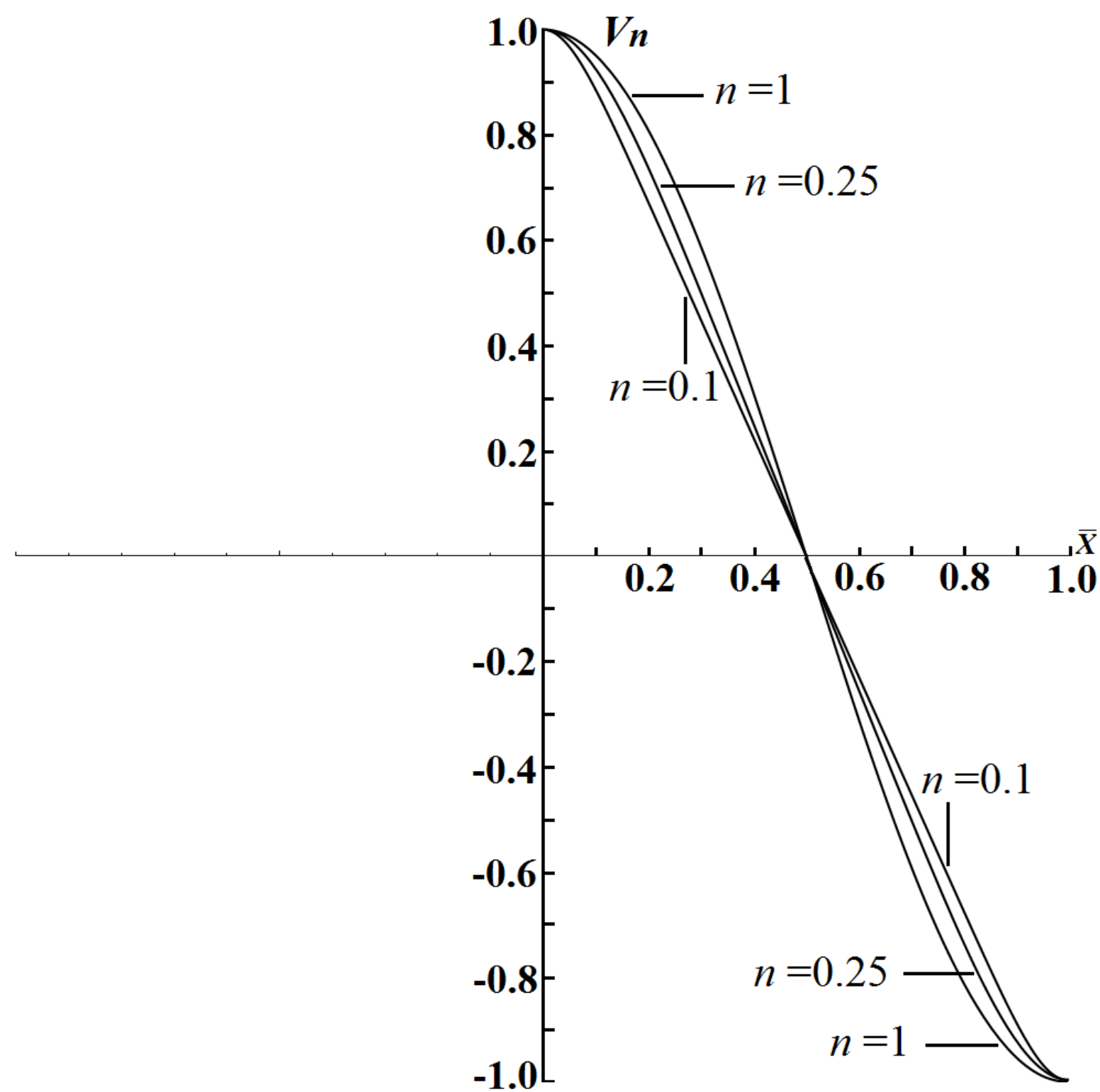}
\caption{Eigenfunction (\ref{eq:v3}) for $n=1,0.25,0.1$.}
\label{fig:slideslideload}
\end{figure}
\end{center}

In the previous two sections, with either PP or PS boundary conditions and convex $u,$ we were able to write a solution $v_n=(u'')^n,\, n\in(0,1)$ to equation (\ref{eq:probscaled}) in terms of generalized trigonometric functions that reduce to their standard trigonometric analogues when $n=1$. In this last setting, we demonstrate how to construct a solution to (\ref{eq:probscaled})-(\ref{eq:bc2})iii based on this simple observation. Let 
\begin{equation}
\label{eq:uv}
v_n(\bar x)=\lvert u^{\prime\prime}(\bar x)\rvert^{n-1}u^{\prime\prime}(\bar x),
\end{equation}
so that $v'_n=n\left|u''\right|^{n-1}u'''.$ Then, the SS boundary conditions (\ref{eq:bc2})iii) imply that $v'_n(0)=v'_n(1)=0$ as long as $u''(0),u''(1)\neq0.$ But notice that for $n=1$ and $\lambda_1=\pi^2,$ $\cos(\pi\bar x)=v_1(\bar x)=u''(\bar x)$ satisfies both of these conditions as well as equation (\ref{eq:probscaled}). Also, recall that when $n=1,$ $\lambda_1=\pi^2$ coincides with the eigenvalue (\ref{eq:firsteig1}), while $\cos(\pi\bar x)=\sin(\pi(1/2-\bar x))$ is simply a translation/reflection of the eigenfunction $\sin(\pi\bar x)$ in (\ref{eq:eigf1}). In this way, to construct a solution for (\ref{eq:probscaled})-(\ref{eq:bc2})iii) with $n\in(0,1)$, we consider, analogue to (\ref{eq:eigf1}), the generalized eigenfunction 
\begin{equation}
\label{eq:v3}
v_n(\bar x)=
\begin{cases}
\frac{2n}{n+1}\sin_n(\pi_n(1/2-\bar x)),\,\,\,\,\,\,\,\,\,&\bar x\in[0,1/2],
\\
-\frac{2n}{n+1}\sin_n(\pi_n(\bar x-1/2)),\,\,\,\,\,\,\,\,\,&\bar x\in[1/2,1].
\end{cases}
\end{equation}
Notice that $v_n(1/2)=0$ and since $\sin_n(\pi_n/2)=\frac{q}{2}=\frac{n+1}{2n},$ we have that $v_n(0)=1$ and $v_n(1)=-1.$ Also,  
\begin{equation}
\label{eq:vsign}
v_n(\bar x)
\begin{cases}
\geq0,\,\,\,\,\,\,\,\,\,&\bar x\in[0,1/2],
\\
\leq0,\,\,\,\,\,\,\,\,\,&\bar x\in[1/2,1].
\end{cases}
\end{equation}
Then, by (\ref{eq:uv}) and (\ref{eq:vsign})i), $v(\bar x)=(u^{\prime\prime}(\bar x))^n$ for $[0,1/2].$ As a result, equation (\ref{eq:probscaled}) reduces to $$\frac{d^2}{d\bar x^2}\left(\frac{2n}{n+1}\sin_n(\pi_n(1/2-\bar x))\right)=-\lambda_n\left(\frac{2n}{n+1}\sin_n(\pi_n(1/2-\bar x))\right)^{\frac{1}{n}},\,\,\,\,\,\,\,\,\,\,\,\bar x\in[0,1/2].$$ Formulas (\ref{eq:dergensin}) and (\ref{eq:dergencos}) then yield (\ref{eq:firsteig1}).
Similarly, (\ref{eq:probscaled}), (\ref{eq:uv}), (\ref{eq:v3})ii) and (\ref{eq:vsign})ii) imply $$\frac{d^2}{d\bar x^2}\left(-\frac{2n}{n+1}\sin_n(\pi_n(\bar x-1/2))\right)=\lambda_n\left(\frac{2n}{n+1}\sin_n(\pi_n(\bar x-1/2))\right)^{\frac{1}{n}},\,\,\,\,\,\,\,\,\,\,\,\bar x\in[1/2,1],$$
and (\ref{eq:firsteig1}) follows from (\ref{eq:dergensin}) and (\ref{eq:dergencos}). We conclude that $P_{cr}(n)$ for the SS boundary conditions is given by (\ref{eq:critload1}).
Finally, $u''$ is given by
\begin{equation}
\label{eq:u3}
u''(\bar x)=
\begin{cases}
\left(\frac{2n}{n+1}\sin_n(\pi_n(1/2-\bar x))\right)^{\frac{1}{n}},\,\,\,\,\,\,\,\,\,&\bar x\in[0,1/2],
\\
-\left(\frac{2n}{n+1}\sin_n(\pi_n(\bar x-1/2))\right)^{\frac{1}{n}},\,\,\,\,\,\,\,\,\,&\bar x\in[1/2,1]
\end{cases}
\end{equation}
and satisfies $u'''(0)=u'''(1)=0.$ Also, as required, $u''$ takes on the nonzero boundary values $u''(0)=1,\, u''(1)=-1$ and, from (\ref{eq:u3}), we see that $u$ has a unique inflection point $\bar x=1/2.$ Finally, formulas for $u$ and $u'$ can be easily obtained from (\ref{eq:dergensin}), (\ref{eq:dergencos}), (\ref{eq:u3}) and $u(0)=u(1)=0.$ See figure \ref{fig:slideslideload} below for a plot of (\ref{eq:v3}).

\section{Connection to Euler- Engesser's Reduced Modulus Critical Buckling Loads}
\label{sec:cp}

Due to the lack of a clear-cut definition of a yield stress or strain for the Hollomon materials, it is somewhat difficult for us to compare our results with the Euler-Engesser's reduced tangent modulus formula $P_{cr}^{(R)}=(\pi/L)^2E_rI$ in general. However, under certain conditions, we can establish a simple relation between the two. Following our discussion in section \ref{sec:introduction}, assume that the neutral axis passes through the centroid of the cross section of the column and replace the Hollomon's power-law (\ref{eq:h}) by
\begin{equation}
\label{eq:ap}
\sigma=
\begin{cases}
E\epsilon,\,\,\,\,\,&|\epsilon|\le {\epsilon}_y,
\\
K|\epsilon|^{n-1}\epsilon,\,\,\,\,\,&|\epsilon|> {\epsilon_y},
\end{cases} 
\end{equation}
where $\epsilon_y$ is an experimentally defined yield strain (see remark \ref{rem:rempp2} below). This means $E=K|\epsilon_y|^{n-1}$ for $|\epsilon|\le {\epsilon}_y$ and the tangent modulus becomes $E_T=nK|\epsilon_y|^{n-1}=nK^\frac{1}{n}|\sigma_y|^{1-\frac{1}{n}}.$ As a result, the corresponding Euler-Engesser's reduced tangent modulus now reads $E_r=\frac{(I_1+nI_2)K|\epsilon_y|^{n-1}}{I}$, and
\begin{equation}
\label{eq:hloadpp}
P_{cr}^{(R)}(n)=\left(\frac{\pi}{L}\right)^2\left(I_1E+I_2nK|\epsilon_y|^{n-1}\right)=\left(\frac{\pi}{L}\right)^2\left(I_1+nI_2\right)|\epsilon_y|^{n-1}K.
\end{equation}
Notice that for the pinned-pinned set of boundary conditions studied in section \ref{subsec:pp}, it follows that (\ref{eq:critload1}) and (\ref{eq:hloadpp}) satisfy 
\begin{equation}
\label{eq:Qpp}
P_{cr}(n)=\frac{2n}{n+1}\left(\frac{\pi_n L^{1-n}}{\pi}\right)^2\left(\frac{I(n)}{I_1+nI_2 }\right)|\epsilon_y|^{1-n}P_{cr}^{(R)}(n),
\end{equation}
and that for $n=1$, $P_{cr}(1)=\frac{I}{I_1+I_2 }P_{cr}^{(R)}(1).$ See the next section for an application of the above formulas to beams with circular or rectangular cross sections. In particular, it can be observed that

$$\frac{P_{cr}^{(R)}(n)}{P_{cr}(n)}=Q(n)=O\left(L^{2(n-1)}\right),$$

\noindent
therefore $\lim_{L\to\infty}\frac{P_{cr}^{(R)}(n)}{P_{cr}(n)}=0$ and $0<Q(n)\leq1$ for all $n\in(0,1]$. This observation also demonstrates that for very long and slender columns, our loads can be significantly less conservative than the Engesser's reduced modulus loads, and would help design more efficient slender columns.

\begin{remark}
\label{rem:rempp2}
The approximation of Hollomon's law by (\ref{eq:ap}) is frequently used in the engineering practice. See, e.g. \cite{Chakrabarty} and \cite{YLing} for details. It is also used in the large commercial finite element package ANSYS\cite{ANSYS}.
\end{remark}

\section{Numerical Solutions}
\label{sec:ex}

In this section, we compute critical loads for Euler-Bernoulli columns of different materials with PP boundary conditions and either circular $\bigcirc$ or rectangular $\square$ cross-sections. Also, these loads are compared to the classical Euler-Engesser's reduced loads. Recall that the generalized, constant moment of inertia $I(n)$ for a beam with cross-section $A$ is given by $I(n)=\int_{A}{\lvert y\rvert^{n+1}dydz}$. Using this formula we find (see appendix) that
\begin{equation}
\label{eq:inertiacircrect}
\begin{split}
I(n)=\left(\frac{2\sqrt{\pi}\,R^{n+3}}{n+3}\right)\frac{\Gamma\left(1+\frac{n}{2}\right)}{\Gamma\left(\frac{3}{2}+\frac{n}{2}\right)}\,\,\,\,\,\,\,\,\,\,\,\bigcirc,\,\,\,\,\,\,\,\,\,\,\,\,\,\,\,\,\,\,\,I(n)=\frac{2h}{n+2}\left(\frac{\omega}{2}\right)^{n+2}\,\,\,\,\,\,\,\,\,\,\square
\end{split}
\end{equation}
for an Euler-Bernoulli column with either circular cross-section $A$ of radius $R>0,$ or a rectangular cross-section $A$ of height $h$ and width $\omega,$ respectively. For PP boundary conditions, the corresponding critical loads are then obtained from $(\ref{eq:critload1})$ and the above generalized moments as
\begin{equation}
\label{eq:critloadpp}
\begin{split}
P_{cr}(n)=\left(\frac{4n\sqrt{\pi}\,\pi_n^2KR^{n+3}}{(n+1)(n+3)L^{2n}}\right)\frac{\Gamma\left(1+\frac{n}{2}\right)}{\Gamma\left(\frac{3}{2}+\frac{n}{2}\right)}\,\,\,\,\,\bigcirc,\,\,\,\,P_{cr}(n)=\frac{nhK\pi_n^2\,\omega^{n+2}}{(n+1)(n+2)(2L^2)^{^{n}}}\,\,\,\,\square.
\end{split}
\end{equation}
Consider Euler-Bernoulli columns with:

\noindent
\textit{\textbf{B1.}}\, Length $L=0.5 m$ and circular cross-section of radius $R=0.01 m$. 

\noindent
\textit{\textbf{B2.}}\, Length $L=0.5 m$ and rectangular cross-section of height $h=0.005 m$ and width $\omega=0.02 m$.

%


In Table \ref{table:materials} below, we compute (\ref{eq:critloadpp}) for several values of $n$ and columns \textit{\textbf{B1}} and \textit{\textbf{B2}} composed of well-known materials. For the sake of comparison, we have included in the Table the corresponding Engesser's reduced loads (\ref{eq:hloadpp}). In evaluating (\ref{eq:hloadpp}), we have used the symmetry of the two cross sections considered here to assume that $I_1=I_2=\frac{1}{2}I$, where we recall that $I=I(1)$. Also, we have used the yield strain $\epsilon_y=0.02$, which is the estimated value for many Hollomon materials. More accurate comparisons can be obtained by using our results once precise values for the yield strain are at hand. Further, using the above on (\ref{eq:Qpp}) yields the comparing ratio

\begin{equation}
\label{eq:ratio}
\begin{split}
Q(n)=\frac{P_{cr}^{(R)}(n)}{P_{cr}(n)}=\left(\frac{\pi(n+1)}{2\pi_n L^{1-n}}\right)^2\left(\frac{I}{I(n)}\right)\frac{|\epsilon_y|^{^{n-1}}}{n}.
\end{split}
\end{equation}

\begin{center}
\begin{figure}[!ht]
\centering
\subfigure{
\includegraphics[scale=0.38]{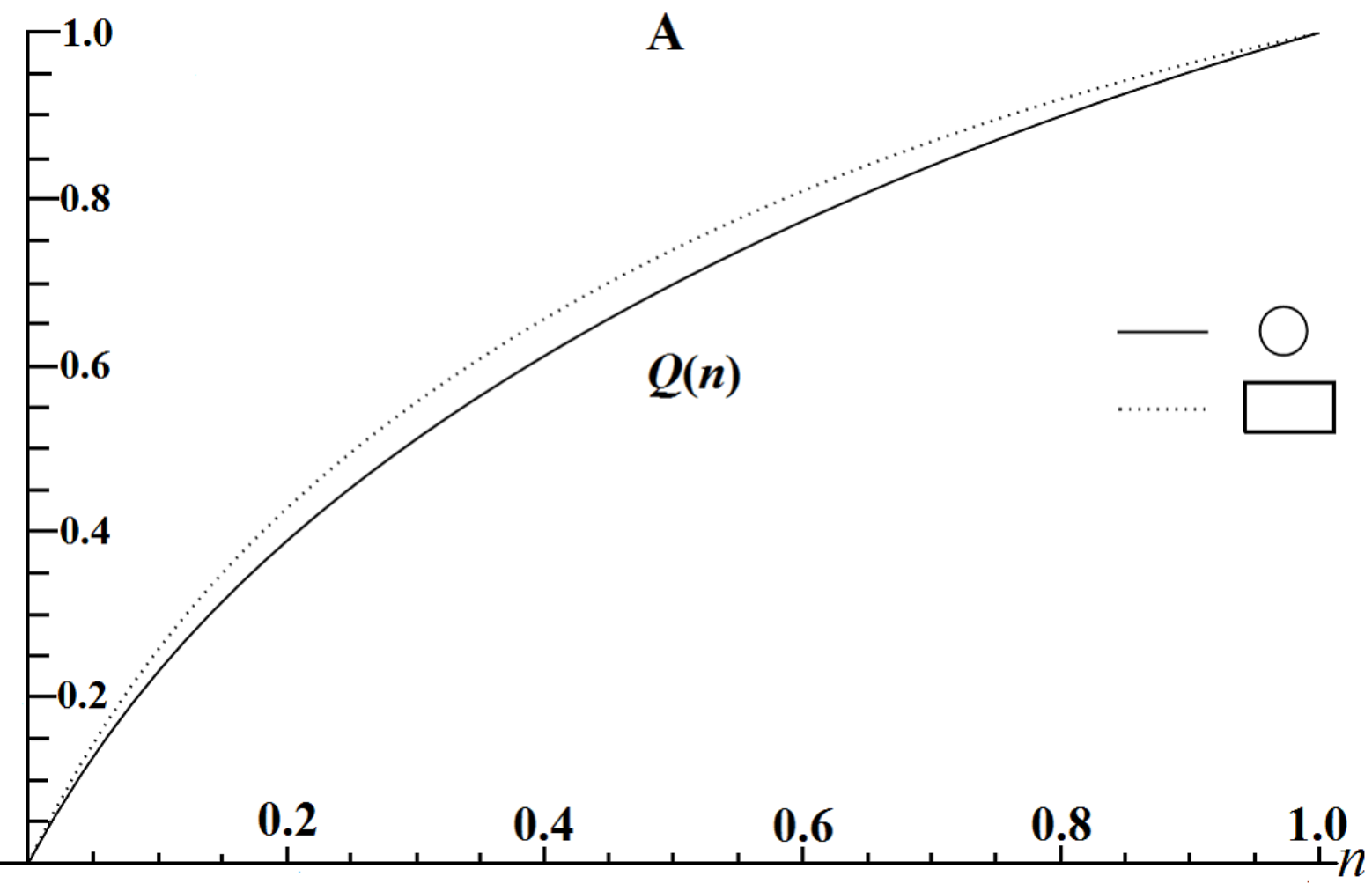}
\label{fig:Qcirclerectangle2}
}
\subfigure{
\includegraphics[scale=0.33]{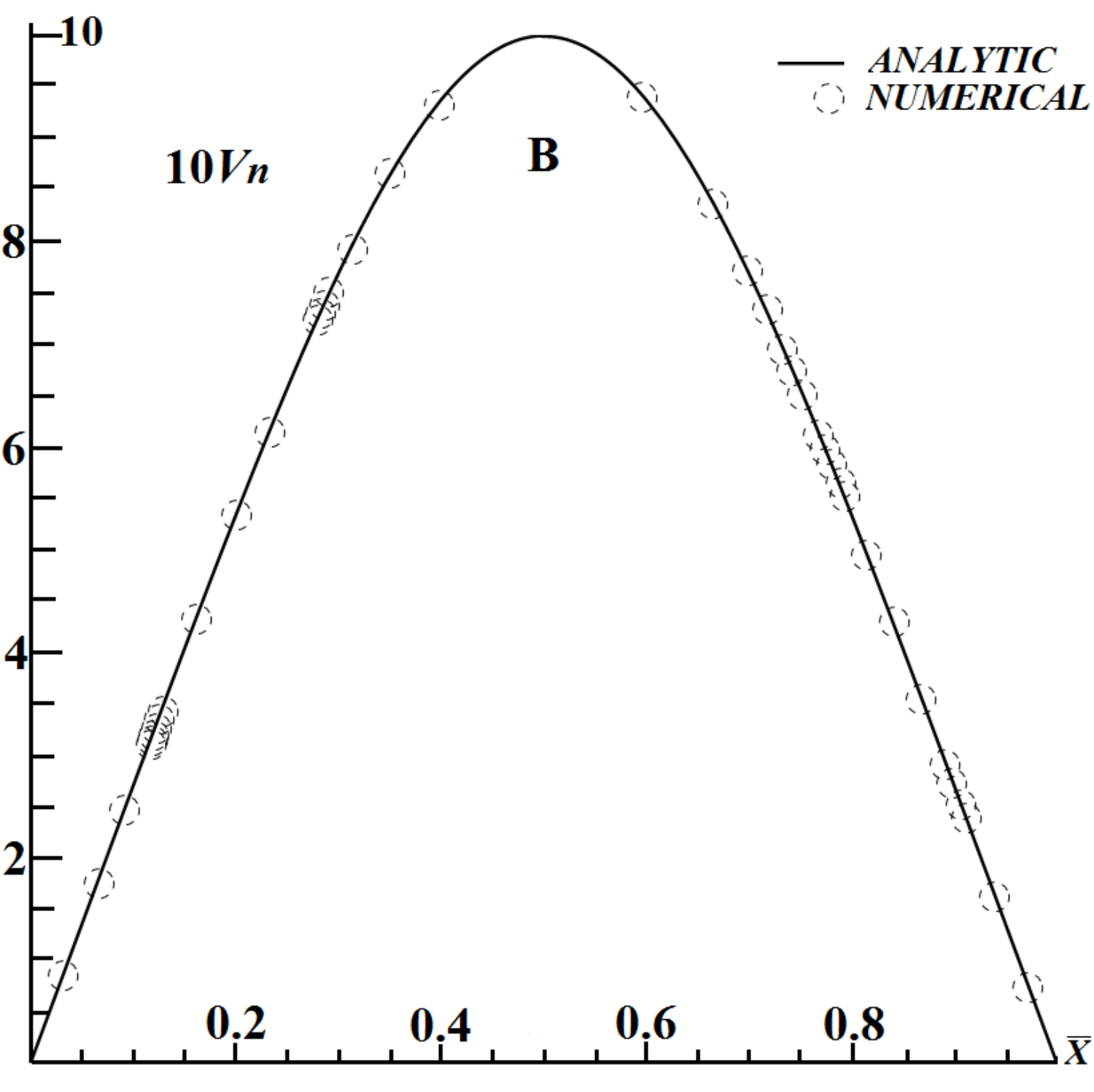}
\label{fig:numericalanalytical2}
}
\label{fig:twofig}
\caption{Figure $A$ depicts the ratio (\ref{eq:ratio}) for $n\in(0,1]$, $\epsilon_y=0.02$ and beams \textit{\textbf{B1}} and \textit{\textbf{B2}} with circular ($\bigcirc$) and rectangular ($\square$) cross-section. Figure $B$ illustrates analytic and numerical solutions to (\ref{eq:secondprob}) for $n=0.4$ and pinned-pinned boundary conditions.}
\end{figure}
\end{center}

Using (\ref{eq:inertiacircrect}) and $\epsilon_y=0.02$, Figure \ref{fig:Qcirclerectangle2} depicts (\ref{eq:ratio}). In agreement with the analytical result, the numerics indicate that our critical buckling load is less conservative than the Euler-Engessers reduced modulus load. 

Lastly, in Figure \ref{fig:numericalanalytical2} we have used the \textsc{ndsolve} \textsc{Mathematica} routine to compute the numerical solution to (\ref{eq:secondprob}) using (\ref{eq:firsteig1}) and $n=0.4$. For comparison purposes, the analytic solution (\ref{eq:eigf1}) is also shown and the error is, approximately, in the order of $ 10^{-4}$. 

We mention that similar type of comparison results for several Euler-type formulae for elastic columns can be found,  e.g., in  \cite{ KardomateasandDancila}.


\begin{table}[ht]
\centering
\caption{Loads (\ref{eq:hloadpp}) and (\ref{eq:critloadpp}) in Kilonewtons (KN) for beams \textit{\textbf{B1}} and \textit{\textbf{B2}} made of well-known materials and various $n$ and bulk modulus $K$ (\cite{Kalpakjian})}
    \begin{tabular}{ | c | c | c | c| c | c | c | c | }
    \hline
       & Material & K(MPa) & $n$ & $P^{(R)}_{cr},\bigcirc$ & $P_{cr},\bigcirc$ & $P^{(R)}_{cr},\square$ & $P_{cr},\square$  \\ \hline
     \textbf{1} & 304 Stainless Steel, ann. & 1275 & 0.45 & 2.46 & 3.76 & 1.05 & 1.5 \\ \hline
     \textbf{2} & 410 Stainless Steel, ann. & 960 & 0.25 & 3.5 & 7.73 & 1.48 & 3 \\ \hline
     \textbf{3} & Aluminum 7075-O & 400 & 0.17 & 1.87 & 5.4 & 0.79 & 2.07 \\ \hline
     \textbf{4} & Aluminum 1100-O & 180 & 0.2 & 0.77 & 1.97 & 0.32 & 0.76 \\ \hline
     \end{tabular}
\label{table:materials}
\end{table}

\begin{center}
\begin{figure}[!ht]
\centering
\includegraphics[scale=0.37]{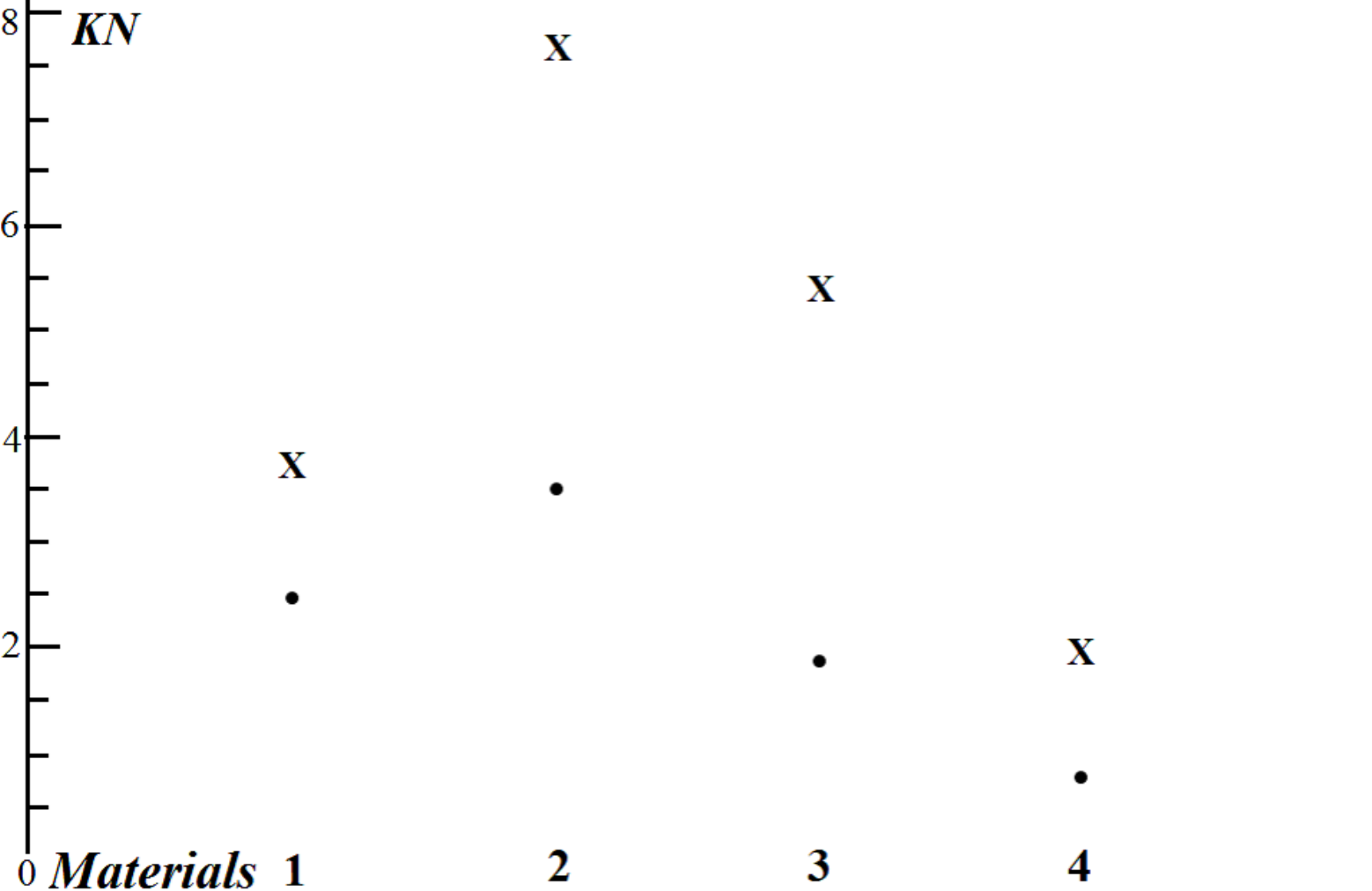}
\includegraphics[scale=0.36]{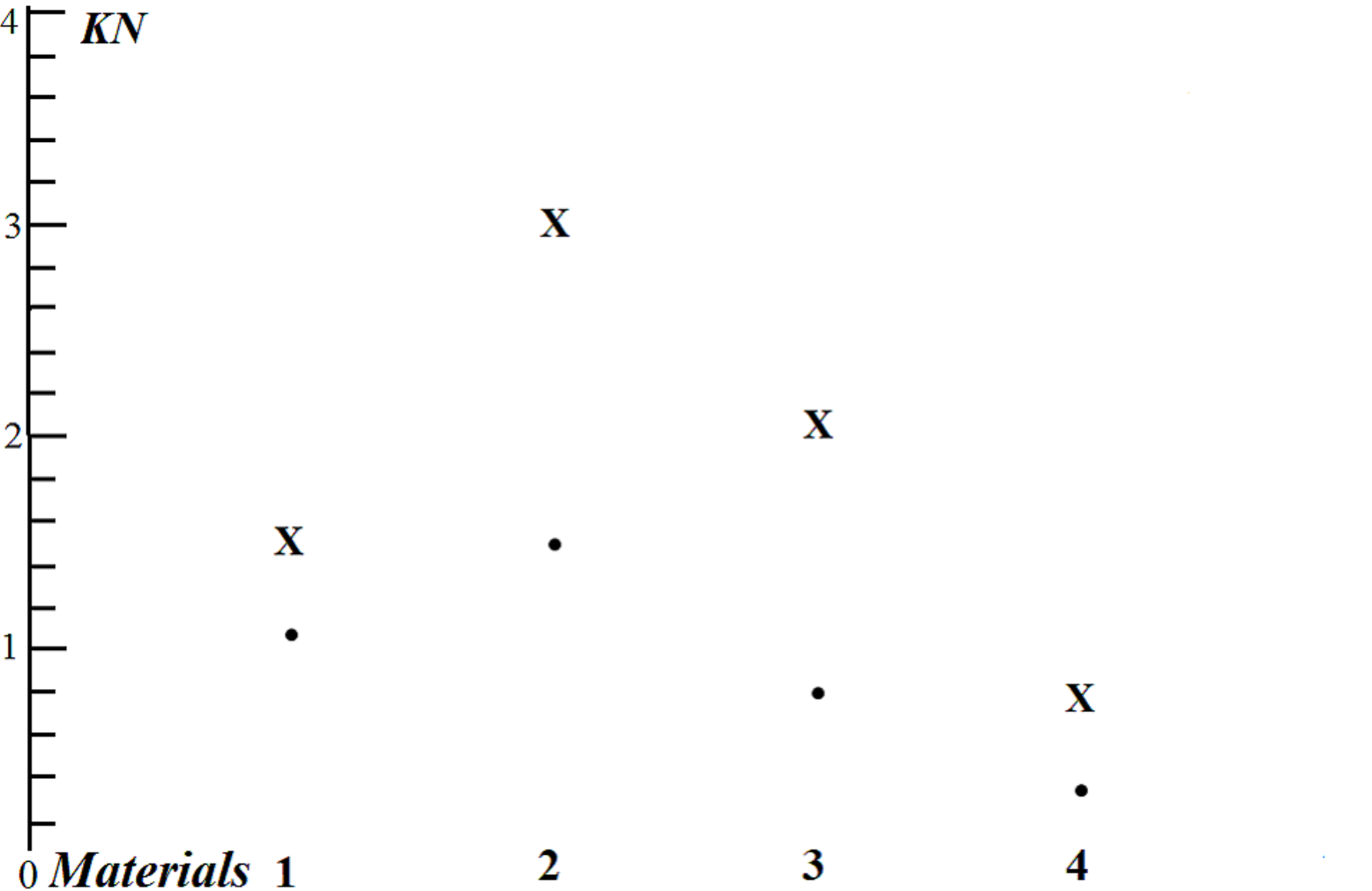}
\caption{For the materials and hardening indexes $n$ given in Table \ref{table:materials}, the above Figures depict, in kilonewtons, both the critical load (\ref{eq:critloadpp}) (marked by \textbf{X}) and the corresponding Engesser's reduced load (\ref{eq:hloadpp}) (solid dots) for beams \textit{\textbf{B1}} and \textit{\textbf{B2}}, respectively.}
\label{fig:exampleloadone}
\end{figure}
\end{center}

\section{Conclusion}
In this paper, we have presented analytic formulas for calculating the global buckling loads of perfect plastic metal columns. The material properties of the columns are assumed to fit the Hollomon's equation.
For simplicity, we considered only pinned-pinned (PP), pinned-slide (PS) and slide-slide (SS) boundary conditions; although similar formulas can be derived for other boundary conditions, and calculations of the critical loads for columns with different cross-sections follow from these formulas. Particularly, for the PP case, we computed the critical loads for columns with either circular or rectangular cross-sections made of several common materials and for some values of $n$. These were then compared to Engesser's reduced modulus critical buckling loads. As demonstrated by the examples, we also showed that for slender columns, our critical loads are significantly less conservative than the reduced modulus loads. The formulae derived in this work can be useful for designing thin or thick columns for a large class of high strength metals modeled by the Hollomon's powerlaw in the plastic range.


\vspace{0.1in}

\textbf{Acknowledgments:} The authors would like to express their gratitude to the referees. Their comments and suggestions led to important revisions and improvements in the contents of this article.

\section*{Appendix - Derivation of (\ref{eq:inertiacircrect})}
\label{sec:app}

For a beam with circular cross-section $A$ of radius $R>0$, we use polar coordinates on $I(n)=\int_{A}{\lvert y\rvert^{n+1}dydz}$ to obtain
\begin{equation}
\label{eq:inertiader}
\begin{split}
I(n)=\int_{A}{\lvert y\rvert^{n+1}dydz}=\int_0^{2\pi}{\int_0^R{r^{n+2}\left|\cos\theta\right|^{n+1}dr}d\theta}=\frac{4R^{n+3}}{n+3}\int_0^{\frac{\pi}{2}}{\left|\cos\theta\right|^{n+1}d\theta}.
\end{split}
\end{equation}
Then, using the standard definition of the gamma function:
\begin{equation}
\label{eq:gamma}
\begin{split}
\Gamma(z)=\int_0^{\infty}{e^{-t}t^{z-1}dt},\,\,\,\,\,\,\,\,\,\,\,\,\,\,Re\, z>0,
\end{split}
\end{equation}
we have the following identity (see for instance \cite{Gamelin1}) 
\begin{equation}
\label{eq:beta}
\begin{split}
\int_0^1{t^{p-1}(1-t)^{s-1}dt}=\frac{\Gamma(p)\Gamma(s)}{\Gamma(p+s)},\,\,\,\,\,\,\,\,\,\,\,\,p,s>0,
\end{split}
\end{equation}
also known as the beta function. Therefore, setting $p=\frac{1}{2}$, $s=1+\frac{n}{2}$ and $t=\sin^2\theta$ into (\ref{eq:beta}) yields
$$\int_0^{\frac{\pi}{2}}{\left|\cos\theta\right|^{n+1}d\theta}=\frac{\sqrt{\pi}\,\Gamma\left(1+\frac{n}{2}\right)}{2\,\Gamma\left(\frac{3}{2}+\frac{n}{2}\right)},$$
which we use on (\ref{eq:inertiader}) to establish formula (\ref{eq:inertiacircrect})i). Since the derivation of formula (\ref{eq:inertiacircrect})ii) for a beam with rectangular cross-section $A$ of height $h$ and width $\omega$ is straightforward, we omit the details.

\end{document}